\documentstyle[aps,preprint]{revtex}

\def\prl{{\sl Phys. Rev. Lett.}\ }
\def\PRL{{\sl Phys. Rev. Lett.}\ }

\def\pr {{\sl Phys. Rev.}\ }

\def\sss                               {
     \scriptscriptstyle                       }
\def\raise                             {
     {\sss \dagger}                            }
\def\spin                            {
     \sigma                             }
\def\k                               {
     k                            }

\def\cu  {
   {\bf c}_{\sss \uparrow}      }
\def\cddag  {
   {\bf c}_{\sss \downarrow}^\raise      }
\def\cudag  {
   {\bf c}_{\sss \uparrow}^\raise      }
\def\kdag{K^\dagger}
\def\e {\epsilon}
\def\cspin                               {
     {\bf d}_{\spin}                   }

\def\ckspin                               {
     {\bf c}_{\k\spin}                   }
\def\cspindag                            {
     {\bf d}_{\spin}^{\raise}              }
\def\ckspindag                            {
     {\bf c}_{\k\spin}^{\raise}              }
\def\nup                              {
     n_{\sss\uparrow}                         }
\def\downa{\downarrow}
\def\upa{\uparrow}
\def\gud{
  G_{\sss \upa\downa}                          }
\def\guu{
  G_{\sss \upa\upa}                          }
\def\grud{
  G^{\,r}_{\sss \upa\downa}                         }
\def\gruu{
  G^{\,r}_{\sss \upa\upa}                         }
\def\grdd{
  G^{\,r}_{\sss \downa\downa}                         }
\def\ndown                            {
     n_{\sss\downarrow}                         }

\def\pauli{
	{\bf \sigma}}
\def\sumspin                         {
     \sum_{\spin}               }
\def\Vzero                         {
     V_{\sss 0}                }
\def\Vso                         {
     V_{\sss {\rm SO}}                }
\def\ezero                       {
     \epsilon_{\sss 0}          }
\def\eone                        {
     \epsilon_{\sss 1}          }
\def\etwo                        {
     \epsilon_{\sss 2}          }
\def\deone                        {
     d\epsilon_{\sss 1}          }
\def\detwo                        {
     d\epsilon_{\sss 2}          }
\def\Gtwo                        {
     \Pi          }
\def\Gtwodelta                        {
     \Pi_{\sss \Delta}          }
\def\kf                          {
     k_{\sss F}   }
\def\dr                          {
     |r - r'|   }
\def\dos                         {
     \nu   }
\def\w                         {
     \omega   }
\def\tauphi                    {
     \tau_{\phi}                        }

\def\tauso                    {
     \tau_{\sss {\rm SO}}                        }

\def\cond                     {
     \sigma_{\sss 0}                        }
\def\lb                     {
     \ell                        }
\def\rhoso                  {
     \rho^{({\sss {\rm SO}})}      }
\def\rhosoq                  {
     \rho^{({\sss {\rm SO}})}_q      }

\def\euu {
   \e_{\sss \upa \upa}             }
\def\eud {
   \e_{\sss \upa \downa}             }
\def\edu {
   \e_{\sss \downa \upa}             }
\def\edd {
   \e_{\sss \downa \downa}             }
\def\Gr {
   G^{\, r}                            }
\def\Ga {
   G^{\, a}                            }
\def\intinf                                 {
    \int_{-\infty}^{\infty}                           }
\def\lso                                    {
    \ell_{\sss {\rm SO} }                        }
\def\kb                                     {
    k_{\sss B}                                   }
\def\tk                                     {
    T_{\sss K}                                   }
\begin{document}

\title{Spin-orbit Scattering  and the Kondo Effect}
\bigskip
\author{Yigal Meir}
\address{Physics Department, University of California, Santa Barbara, CA 93106}
\author{Ned S. Wingreen}
\address{NEC Research Institute, 4 Independence Way, Princeton, NJ 08540}
\bigskip
\maketitle
\begin{abstract}
The effects of spin-orbit scattering of conduction electrons
in the Kondo regime are investigated
theoretically. It is shown that due to time-reversal symmetry,
spin-orbit scattering does not suppress the Kondo effect, even though it
breaks spin-rotational symmetry, in full agreement with experiment.
An orbital magnetic field, which breaks
time-reversal symmetry, leads to an effective Zeeman splitting, which
can be probed in transport measurements. It is shown that,
similar to weak-localization, this effect has anomalous magnetic field
and temperature dependence.
\end{abstract}
\vfill
\bigskip
\newpage
The profusion of works on the Kondo effect
in the last thirty years has led to a good
understanding of the strongly correlated state\cite{hewson}, with the possible
exception of systems of reduced dimensionality \cite{bergmann,giordano}.
As the temperature is lowered the electron gas screens the isolated
impurity spin, leading to enhanced scattering on the Fermi surface.
Elastic impurities do not change this picture, and their effect can be
absorbed into renormalizing the Kondo temperature \cite{fischer,kotliar}.
Spin scatterers, on the other hand, are expected to suppress the effect,
as the electrons lose their spin-memory after traveling the spin-scattering
length.

Recently, Bergmann \cite{bergman} has demonstrated in an elegant
experiment that weak-localization effects can be used to study the
effectiveness of the Kondo screening of  magnetic impurities. By
measuring the magnetoconductance for various systems, he was able
to identify the amount of magnetic scattering and, consequently,
the screening of the magnetic impurities. Surprisingly, it was found
that adding a large number of spin-orbit scatterers into the sample
(such that the magnetoconductance changes sign due to the weak-antilocalization
phenomenon) does not change the magnetic scattering at all. Accordingly,
the spin-orbit scattering, even though it breaks the spin-rotational
symmetry of the system, does not suppress the Kondo effect.
Several other groups have also reported the observation of the Kondo effect
in the presence of strong spin-orbit scattering \cite{experiment}.

In this Letter we discuss the effects of spin-orbit scattering in the
Kondo regime. It is shown that because of time-reversal symmetry, the
spin-orbit scatterers play the same role as elastic, non-magnetic
impurities. As the spin-orbit scattering rate  is usually much
 smaller than the elastic scattering rate,
 we expect there will be no observable change
in the Kondo temperature and hence in the Kondo screening of the
magnetic impurities due to  spin-orbit scattering, in full
agreement with the experiment\cite{bergman} .
Interestingly, however, the application of a magnetic field leads to the
breaking of time-reversal symmetry, and consequently, suppresses
the Kondo effect. We calculate the effective Zeeman splitting
resulting from this orbital magnetic field,
 and discuss its experimental implications.

The resonant scattering of the electrons near the Fermi energy can be traced
back to the divergence of the self-energy in a perturbation expansion,
either in  the coupling between the local spin and the electronic spin
 in the Kondo s-d Hamiltonian, or in the
hopping in the Anderson model. Both kinds of self-energies involve multiple
scattering of the conduction electrons by the local impurity and accordingly
involve those electrons only through the propagator $G(r,r;t)$,
from the impurity position, $r$, back to the impurity position.  If the
 electron spin is rotated randomly during that propagation, the correlations
between consecutive scattering events are
lost and the Kondo effect is suppressed.
Here we  prove that when time--reversal symmetry is obeyed,
the propagator $G(r,r;t)$ is diagonal in spin-space. Accordingly, even in the
presence of spin-orbit scattering the Kondo effect persists. For example,
\begin{eqnarray}
<\cu(r,t)\cddag(r,0)>\ \
&=&  \ \ \displaystyle{1\over Z}\ \displaystyle{\sum_{n,m}} \,
e^{-\beta E_m + i(E_m-E_n)t}
<\Psi_m|\cu(r,0)|\Psi_n>\,<\Psi_n|\cddag(r,0)|\Psi_m>  \nonumber \\
=  \ \ \ \  \,
\displaystyle{1\over Z}\ &\displaystyle{\sum_{n,m}}\,&
e^{-\beta E_m + i(E_m-E_n) t}
<\Psi_m|\kdag\cu(r,0)K|\Psi_n>\,
<\Psi_n|\kdag\cddag(r,0)K|\Psi_m> \nonumber \\
=\   -\ \displaystyle{1\over Z}
\ &\displaystyle{\sum_{n,m}}\,& e^{-\beta E_m + i(E_m-E_n) t}
<\Psi_n|\cddag(r,0)|\Psi_m>\,
<\Psi_m|\cu(r,0)|\Psi_n>  \nonumber \\
=  \ \ \ \ \ \,\, - &<&\cu(r,t)\cddag(r,0)>\ \,\ \, = \ \ \ 0 \ \ \ ,
\label{tri}
\end{eqnarray}
where  ${\bf c}_\sigma(r,t)$ (${\bf c}^\raise_\sigma(r,t)$) annihilates
(creates) a conduction electron of spin $\sigma$
at position $r$ and time $t$, and
$K$ is the time-reversal operator.
In the above the propagator was expressed in terms of the exact
many-body eigenfunctions of the system, $\Psi_n$, with energies
$E_n$, and we have used the identities
$<\Psi_m|\kdag{\bf c}_{\sss \upa}(r,0)K|\Psi_n>\, =$\linebreak
$ -\!<\Psi_n|\cddag(r,0)|\Psi_m>$ and
$<\Psi_n|\kdag\cddag(r,0)K|\Psi_m>\, =\, <\Psi_m|\cu(r,0)|\Psi_n>$.
In (\ref{tri}) the chemical potential was taken as zero, for convenience.
The same procedure can be applied to show that $<\cddag(r,0)\cu(r,t)>$,
and, consequently, the retarded Green function,
$\grud(r,r;t)\equiv
-i\theta(t)<\left[\cddag(r,0)\cu(r,t)+\cu(r,t)\cddag(r,0)\right]>$, are
also identically zero, and that $\gruu(r,r;t)=\grdd(r,r;t)$.
Note that the above results apply for any interacting system
satisfying time-reversal symmetry, even including inelastic scattering.

Let us give another, more transparent argument why an electron always returns
with the same spin. Consider a general closed path from point $r$ to itself
(e.g. the trajectory in Fig. 1(a)). Such a path can be schematically
represented
by the left trajectory in Fig. 1(b).
The electron spin is rotated due to spin-orbit scatterers along the path.
Due to
time-reversal symmetry the spin-scattering matrix along the path can be written
as
\cite{friedel,meir}
$\bf{S} = \pmatrix{\alpha &\beta \cr -\beta^*&\alpha^*}$.
The electron can also follow the time-reversed
trajectory (the right trajectory in Fig 1(b)),
where all the scatterers are met in opposite order, which gives rise
to the rotation matrix $\bf{S^{\dagger}}$. Since both trajectories have exactly
the same weight, one can add them up, leading to a matrix proportional to the
unity matrix. Thus it is
the destructive interference between time-reversed paths that leads to the
vanishing of the off-diagonal terms \cite{fnote}.

The above argument suggests that this picture will change dramatically in the
 presence of a magnetic field, which
breaks  time-reversal symmetry. In the absence of spin-orbit scattering
a magnetic field suppresses the Kondo effect  through the Zeeman splitting
 of the impurity state. The peak in the impurity density of states  moves
away from the Fermi energy by the Zeeman splitting \cite{us}. Once the
splitting is larger than the Kondo temperature, the
ground state of the impurity is polarized, suppressing the
resonant Kondo scattering at the Fermi surface.
 The split peaks can still be probed, though, via nonlinear transport
 measurements, where they produce split peaks in the
differential $I-V$ characteristics \cite{us,appelbaum}.

In the presence of spin-orbit scattering an orbital magnetic field leads to
similar effects, as it destroys the exact cancellation of the contributions
of the time-reversed paths to the off-diagonal propagator. Thus
 an electron may return
to the impurity position with a rotated spin, mixing the two spin directions
and giving rise to an effective Zeeman splitting. To calculate this effect we
consider specifically
an Anderson Hamiltonian in the presence
of spin-orbit scattering and magnetic field,
\begin{equation}
 {\cal H} =
   \ezero \,\displaystyle{\sumspin}   \cspindag \cspin  + U \nup \ndown
  + \displaystyle{\sum_{\spin}}
        \,[\Vzero {\bf c}^{\raise}_{\sigma}(0) \cspin + {\rm H.c.}]
  + {\cal H}_{\rm el}   \ \ .
\label{H}
\end{equation}
The operators $\cspindag$ create a local electron on the impurity; the
second term describes the impurity on-site repulsion
($n_{\sigma}\equiv\cspindag \cspin$), while the third term describes the
hopping
between the impurity (positioned at $r=0$)
and the electron gas. The conduction electron Hamiltonian is given by
\begin{equation}
 {\cal H}_{\rm el}=
   {1\over{2m}}\displaystyle{\sum_{k,\spin}}
 \  (\hbar k-{e{\bf A}\over{c}})^2
 \ \ckspindag \ckspin
		\,+\,  \displaystyle{V\sum_{q,p}}\,\rho_q
	{\bf c}_{p+q \spin}^{\raise} {\bf c}_{p \spin} +
	 i\Vso\, \displaystyle{\sum_{q,p}} \, \rhosoq
   {\bf q}\times{\bf p}\cdot
 ({\bf c}_{p+q\,\spin'}^{\raise}\,  \pauli_{\spin'\spin}\,
    {\bf c}_{p\, \spin}),
\label{Hel}
\end{equation}
where $\ckspin$ is the Fourier transform of
${\bf c}_\sigma(r)$,
$\bf A$ is the electromagnetic potential, and $\rho_q$ and $\rhosoq$
are the densities of the elastic scatterers and spin-orbit scatterers,
respectively.

The presence of both spin-orbit
scattering and magnetic field leads to anisotropy in spin-space
of the conduction-electron density of states. Through the hopping,
$V$, this anisotropy lifts
the degeneracy between the spin states of the impurity. The
resulting effective Zeeman splitting is obtained by diagonalizing
the impurity Hamiltonian
\begin{equation}
{\cal H}_{\rm imp} = \pmatrix{\ezero + \delta \euu & \delta \eud \cr
   \delta \edu & \ezero + \delta \edd}.
\label{himp}
\end{equation}
The energy splitting, $\Delta$,  between the eigendirections of
spin is given by
\begin{equation}
 \Delta^2 \equiv (\e_1-\e_2)^2 =
(\delta \euu - \delta \edd)^2 + 4|\delta \eud|^2,
\label{energy}
\end{equation}
where
to lowest order in the hopping (for
$U \rightarrow \infty$),
\begin{equation}
\delta \e_{\sigma \sigma'} = \displaystyle{
    { {i \Vzero^2} \over {2 \pi} } \intinf d\e
 \, {  {1 - f(\e)} \over {\ezero - \e} }
 \, [\Gr_{\sigma \sigma'}(r,r,\e) - \Ga_{\sigma \sigma'}(r,r,\e)]   }.
\label{deltae}
\end{equation}
The energy shift in Eq. (\ref{deltae}) is due to processes in which
the impurity electron hops to an unoccupied state in the conduction
band, propagates as a conduction electron, and finally hops back to
the impurity, possibly with a rotated spin.
Combining Eqs. (\ref{energy}) and (\ref{deltae}), we find
that the effective Zeeman splitting, $\Delta$, is given by \cite{kondo}
\begin{equation}
\Delta^2\ \  = \ \ {\Vzero^4\over{(2\pi)^2}} \int\!\!\int\! \deone\detwo
{ {[1-f(\e_1)][1-f(\e_2)]}
\over {(\eone - \ezero)(\etwo - \ezero)}    }
\  \Gtwodelta(\eone,\etwo) \ ,
\label{splitting}
\end{equation}
where
\begin{equation}
\Gtwodelta(\eone,\etwo) \,=
\, \Gtwo_{\sss \upa\upa\upa\upa}(r,r;\eone,\etwo)
+  \Gtwo_{\sss \downa\downa\downa\downa}(r,r;\eone,\etwo)
-  2\Gtwo_{\sss \downa\downa\upa\upa}(r,r;\eone,\etwo)
+  4\Gtwo_{\sss \downa\upa\upa\downa}(r,r;\eone,\etwo) ,
\label{Pidef}
\end{equation}
and
\begin{equation}
   \Gtwo_{\alpha\beta\gamma\delta}(r,r;\eone,\etwo) = - \,
   [\Gr_{\delta\gamma}(r,r;\etwo) - \Ga_{\delta\gamma}(r,r;\etwo)]
   [\Gr_{\beta\alpha}(r,r;\eone)  - \Ga_{\beta\alpha}(r,r;\eone)] .
\label{Gtwo}
\end{equation}
In the absence of magnetic field,
$\Gtwodelta$ is explicitly zero and there is therefore no
effective Zeeman splitting of the impurity.

 The magnetic-field dependence of the splitting
is determined by the  magnetic-field dependence of $\Gtwodelta$, which,
after averaging over disorder,
is determined by the cooperon diagram (Fig. 1(c)).
 The cooperon  diagram has been calculated
in the context of weak localization theory  \cite{patrick,kawabata,boris},
 and the difference between $\Gtwodelta$ at finite field and its zero field
value is given in three dimensions by
\begin{equation}
 <\Gtwodelta^{^{(3d)}}(\w)>  \ =
  \ \displaystyle{
{{3 \dos} \over {\hbar D(1 - i\w \tau)^2}\lb }}
   \ \displaystyle{ \left\{\,
        \,F\left[{ {\lb^2} \over {4 \hbar D} }
\left(-i\w + {\hbar \over \tauphi} \right) \right]
   -
\,F\left[{ {\lb^2} \over {4 \hbar D} }
\left(-i\w +  {4\hbar \over {3\tauso}} + {\hbar \over \tauphi} \right) \right]
 \right\} }
\label{Pi3d}
\end{equation}
where
$\w = \eone - \etwo$, and
\begin{equation}
F(\zeta) = \sum_{{\sss N} = 0}^{\infty} \,\left[
   \  2\,\Bigl(\sqrt{ N + 1 + \zeta}
       - \sqrt{N + \zeta}\,\Bigr)
   \ -\  { 1 \over {\sqrt{ N + 1/2 + \zeta}} }\ \right]  \ \ \ .
\label{Fdef}
\end{equation}
In two dimensions, we find
\begin{eqnarray}
 <\Gtwodelta^{^{(2d)}}(\w)>  \ =
  \  &\displaystyle{
{{3 \dos} \over {\hbar D(1 - i\w \tau)^2} }
  \ \left\{\  \Psi\left[\, {1\over 2}
+ { {\lb^2}\over{4\hbar D}}\left(-i\w + {\hbar\over\tauphi}\right)\right]
   \right. }
 \nonumber \\[10pt]
&\displaystyle{\left.
   \ - \  \Psi\left[\, {1\over 2}
     + { {\lb^2}\over{4\hbar D}}\left(-i\w +
    {{4\hbar} \over {3\tauso}}
    + {\hbar\over\tauphi}\right) \right]
\ -\ \log\left(
{{-i\w+\hbar/\tauphi}\over{-i\w+4\hbar/3\tauso+\hbar/\tauphi}}
\right)
\ \right\} }\  \ \ .
\label{Pi2d}
\end{eqnarray}
In the above $\dos$ is the single-spin conduction-electron density of states,
$\lb$ is the magnetic length $\sqrt{\hbar c/eH}$, $D$ is the diffusion
constant,
$\tauso$ is the spin-orbit scattering time
 ($=\hbar/[2\pi\dos\rhoso_{q=0}\Vso^2\overline{(p\times q)^2}]$),
 $\tauphi$ is the phase breaking time, and $\Psi$ is the Digamma function.
In deriving those results it was assumed that $\tau\ll\tauso,\tauphi$,
where $\tau=\hbar/(2\pi\dos\rho_{q=0} V^2)$ is the elastic lifetime.
By inspection, only the real
parts of expressions (\ref{Pi3d}) and (\ref{Pi2d}) contribute to the
effective Zeeman splitting given by Eq. (\ref{splitting}).

Since the magnetic-field dependence of the cooperon diagram
 also determines the magnetoconductance
in the weakly localized regime, we can deduce the magnetic field dependence
of the splitting from the weak-localization magnetoconductance.
 Thus we expect the splitting to be linear
in small magnetic fields, crossing over  at high fields to $\sqrt{\log(H)}$
in two dimensions and to $H^{1/4}$ in three dimensions. To see the amplitude
of the effect we expand the $\Pi$'s in small magnetic field and find
\begin{eqnarray}
<\Delta^2\,>  \ \ \   &\stackrel{\rm {\sss 3d}}{=}&\,\ \ \
\displaystyle {2\pi\, \left[ { m \over {m^*} }
   { { \mu_{\sss B} H} \over{4\log (W/\kb\tk)}}\right]^2
{ {  \cond} \over {e^2/h} }\, {1\over{\kf^2 \lso}}}  \nonumber\\[10pt]
  &\stackrel{\rm {\sss 2d}}{=}&\, \ \ \ \displaystyle {{2\over \pi}\,
\left[{m \over {m^*}}
{ { \mu_{\sss B} H}
   \over{4\log (W/\kb\tk)}}\right]^2{ {  \cond} \over {e^2/h} }\,}
\displaystyle { \log \left( { {3\tauphi} \over {4\tauso} }\right) } ,
\label{Deltaresult}
\end{eqnarray}
where $\mu_B=e\hbar/2mc$ is the Bohr magneton, $W$ is the band width of the
 conduction electrons, $m^*$ is their effective mass,
 $\cond$ is the conductivity (conductance in
two dimensions),
and $\lso= \sqrt{D\tauso}$ is the spin-orbit length.
For quasi two-dimensional systems (where the thickness
 of the sample, $d$, is larger than one-half the  Fermi wavelength)
 the two-dimensional
result has to be divided by the square of the number of subbands, $\kf d/\pi$.
To obtain Eq. (\ref{Deltaresult}) it was assumed that the depth
of the impurity level, $\mu - \ezero$, is much larger than the
energy broadening due to elastic scattering, $\hbar/\tau$,
and that $\tauphi \gg \tauso$.
The result indicates that the splitting in a finite magnetic field
increases the more conductive the sample is. In two dimensions it is also
predicted that the splitting depends on the temperature logarithmically
through the inelastic lifetime.
We thus predict that nonlinear measurements will reveal a temperature
dependent splitting, with anomalous magnetic field dependence at high
fields.  In typical metallic samples, three dimensional or quasi
 two-dimensional,  the amplitude of the effect is only a small
 fraction of the usual Zeeman splitting for typical
experimental values \cite{experiment}. This explains why the Kondo
effect has been observed in experiments with strong spin-orbit
scattering even in the presence of a magnetic field \cite{bergman,experiment}.
On the other hand, the effect is expected to be
much larger in two-dimensional semiconductor systems, because  of the
reduced dimensionality and because of the higher mobility.
 The Kondo effect has indeed been observed in dilute magnetic
 semiconductors\cite{dms,kondosemi}, while spin-orbit scattering in the
weakly localized regime has been systematically investigated in several
semiconductor compounds \cite {sosemi}. In fact, spin-orbit scattering in the
weakly localized regime has been  recently reported in a dilute magnetic
semiconductor \cite{both}. With the experimentally measured parameters
for a two-dimensional electron gas in
Hg$_{0.79}$Cd$_{0.19}$Mn$_{0.02}$Te
reported in Dietl {\it et al.} \cite {both},
where both spin-orbit and magnetic scattering have been observed,
the impurity splitting due to the orbital magnetic field
is an order of magnitude larger than
the usual Zeeman splitting.  In addition, with
the progress in building heterostructures
involving diluted magnetic semiconductors \cite{jeremy}, it should be
feasible to systematically
check the predictions of our theory.

To conclude, we have studied in detail spin-orbit scattering in the
Kondo regime. It was shown that due to time-reversal symmetry,
spin-orbit scattering, even though it breaks spin-rotation symmetry, does not
suppress the Kondo effect.
This explains the surprising results of Bergmann \cite{bergman}
on thin films containing both Kondo impurities and spin-orbit
scatterers.
We find that in a finite magnetic field,
which breaks time-reversal symmetry, spin-orbit scattering leads to
an effective Zeeman splitting, with anomalous magnetic field
and temperature dependence, similar to the magnetoconductance in
the weakly localized regime. It is hoped that this work will stimulate
further experiments to explore these effects.

We acknowledge discussions with Boris Altshuler.
Work at U.C.S.B. was supported by NSF Grant No.
NSF-DMR-9308011, by the NSF Science and Technology Center
for Quantized Electronic Structures, Grant No. DMR 91-20007,
and by NSF, ONR, and ARO at the Center for Free Electron Laser
Studies.

\bigskip
\bigskip
\bigskip
\bigskip
\bigskip
\bigskip
\bigskip
\bigskip
Figure Caption

\medskip
\parindent=0pt
\hoffset -1.0 truecm
(a) Schematic path contributing to the local density of
states at a Kondo impurity.
Even with spin-orbit scattering by impurities (the X's),
the equal weighting of time-reversed paths (b)  guarantees
a diagonal  single-particle propagator
$G_{\beta\alpha}(r,r)$, which preserves the
Kondo effect.
 (c) The cooperon
contribution,\linebreak
$< G^a_{\delta\gamma}(r,r;\etwo)
            G^r_{\beta\alpha}(r,r;\eone)>,$
to the disorder-averaged two-particle propagator, \linebreak
$<\Gtwo_{\alpha \beta \gamma \delta}(r,r;\eone, \etwo)>$.


\begin{references}
\bibitem[10]{fnote}In an actual physical system an electron
may hop from the impurity to some point,
 travel around and then hop back onto the impurity from a nearby point.
 As in this case the time-reversed trajectories
no longer go between the same points, they accordingly acquire an additional
relative phase, $k_F|r-r'|$, where $k_F$ is the Fermi wave length.
Indeed one can show that
$ {{<|\gud^r(r,r';t)|^2>}/
 {<|\guu^r(r,r;t)|^2>}}\,
  =
  \,1\, -\, { {\sin^2(\kf\dr)} / {(\kf\dr)^2} } \, .$
As the spatial extent of the impurity (d or f-level) wavefunction is much
smaller
than $\hbar/k_F$, we expect that a given electron will always hop into the
impurity
with the same spin, even in the presence of  spin-orbit scattering.
\end{references}
\end{document}